\documentclass{basi}
\usepackage{graphicx}
\newcommand{\beq}{\begin{equation}}
\newcommand{\eeq}{\end{equation}}
\newcommand{\beqa}{\begin{eqnarray}}
\newcommand{\eeqa}{\end{eqnarray}}
\begin{document}
\title[The non-linear theory of a warped $\beta$-disc]{The non-linear theory
of a warped accretion disc with the $\beta$-viscosity
prescription}
\author[J. Ghanbari , M. Shadmehri and F. Salehi]
       {Jamshid
       Ghanbari\thanks{e-mail:ghanbari@ferdowsi.um.ac.ir},
       Mohsen
       Shadmehri\thanks{e-mail:mshadmehri@science1.um.ac.ir} and
        Fatemeh Salehi\thanks{e-mail:fsalehi@wali.um.ac.ir}\\
        Department of Physics, School of Sciences, Ferdowsi
University of Mashhad, Mashhad, Iran}

\maketitle \label{firstpage}
\begin{abstract}
We study the nonlinear dynamics of a warped or twisted accretion
disc, in which the viscosity coefficients are assumed to be
locally proportional to the rotational velocity ($\beta-$
prescription). Using asymptotic methods for thin discs, dynamical
equations of the disc are obtained in a warped spherical polar
coordinates. These  equations  are solved by the method of the
separation of the variables. This analysis constitutes an
analogous study of the nonlinear theory of an alpha model warped
disc which has been studied by Ogilvie (1999). We have compared
our results with Ogilvie's analysis. The dynamical behaviors of
these models have also been discussed. Our results show that
different viscosity prescriptions and magnitudes ($\alpha$ and
$\beta$ prescriptions) affect dynamics of a warped accretion disc.
Therefore it can be important in determining the viscosity law
even for a warped disc.
\end{abstract}
\begin{keywords}
accretion, accretion discs -- hydrodynamics
\end{keywords}
\section{Introduction} The development of studies about accretion discs
during present century is an illustration of the grow of interest
of researches. In more recent times a good deal of attention has
been devoted to studies of the warped accretion discs. This kind
of discs has been observed in a wide variety of astronomical
objects from young stellar objects, X-ray binary stars to active
galactic nuclei. For example, the stability of the super orbital
periodicity in the neutron star XRBs Cyg X-2, LMC X-4 and Her X-1
(Clarkson et al. 2003) is attributed to a warped disc. The
outflows and illumination patterns from the central engine of a
Seyfert Galaxy (Greenhill et al. 2003) provide direct reasoning to
warped accretion discs. Also, observations of FUSE (Far
Ultraviolet Spectroscopic Explorer) and HST (Hubble Space
Telescope) ultraviolet  of the low-inclination, nova-like
Cataclysmic variable RW Sex presented evidences of warped discs
(Prinja et al. 2003). The precession of warped discs in magnetized
stars (T Tauri stars, white dwarf or neutron stars) discussed by
Pfeiffer \& Lai (2004) that can be concluded from magnetic torques
due to the interaction between the stellar field and the induced
electric currents in the disc. A photometric study of the SW Sex
star, PX And (Boffin et al. 2003) reveals a precessing disc
possibly warped. Many different systems, including young stellar
objects and X-ray binaries, display  a different precession period
or radiation flux variabilities that it may be concluded from a
warped accretion disc. Also,  Her X-1 (Tananbaum et al. 1972; Katz
1973; Roberts 1974; Still \& Boyd 2004) and the hard X-ray
component of the micro quasar GRS1915+105 (Rau, Greiner \&
McCollough 2003) show another typical behavior of warped discs.

During recent years, several driving mechanisms for disc warps
have been suggested. A warp may be induced through instabilities
that may be due to viscous torque (Pringle 1992), bending waves
and viscous torques (Ogilvie 1999), resonant tidal interactions
(Lubow 1992; Lubow \& Ogilvie 2000); irradiation-driven wind
torques (Schandl \& Meyer 1994); radiation torques (Pringle 1996);
wind torques via Kelvin-Helmholts instability (Quillen 2001);
magnetic torques (Lai 1999) and magnetic-induced electric current
torques (Lai 2003).

Most theoretical studies of the accretion discs are based on the
concept of a real fluid and the equations of magnetohydrodynamics.
Adjacent layers of a moving real fluid experience tangential
forces (shearing stresses) as well as normal forces (pressures).
It seems that the viscous forces are the main physical agent in
any accretion disc theory. Therefore,  the viscous forces are
parameterized. A way of doing this is through the dimensionless
parameter $\alpha$ introduced by Shakura \& Sunyaev (1973). It
describes angular momentum transport in accretion discs. The
mechanisms of the angular momentum transport can be described by
the magnetic or the hydrodynamic instabilities. There is another
prescription  through the dimensionless parameter $\beta$
introduced by Lynden-Bell \& Pringle (1974), in which the
viscosity coefficient is proportional to the rotational velocity.
Analyzing the turbulent flows between the coaxial cylinders,
Richard \& Zahn (1999) investigated $\beta$- prescription for
turbulent viscosity and found that the parameter $\beta$ is of the
order of $10^{-5}$. Furthermore, in their model, the viscosity is
very small comparing with the viscosity in $\alpha$- prescription.
Consequently, the forces due to viscous friction are very small
compared with the remaining forces (gravity and pressure forces).
So, the Reynolds numbers are very large, because of very low
viscosity of the fluid. Accretion disc model with $\beta$-
prescription has been studied by Hure, Richard \& Zahn (2001).
They argued  that the beta model for viscosity prescription is
applicable for analyzing the steady state structure of the
Keplerian accretion discs and it yields somewhat different results
compared to the classical $\alpha$-viscosity introduced by Shakura
\& Sunyaev (1973). Study on hydrodynamic viscosity and
selfgravitation in non-warped accretion discs ($\beta$ model) has
been done by Duschl et al.(2000). They showed that $\beta$-discs
can explain the observed spectra of protoplanetary discs and yield
a natural solution to an inconsistency in the $\alpha$-disc models
if the mass of the disc is large enough for self-gravity to play a
role and in the limit of low mass, hydrodynamic turbulence will
result by $\alpha$ model. Turbulence induced by the horizontal and
vertical shear has been studied by Mathis et al. (2004). They have
presented a new prescription, the $\beta$-viscosity, for the
horizontal component of the turbulent diffusivity due to the
differential rotation in latitude. They generalized $\beta$
prescription (Richard \& Zahn 1999) in the stationery limit,
advection and diffusion balance each other. Because that
prescription (Richard \& Zahn 1999) has been established in the
case of maximum differential rotation and so its validity must be
verified to milder shear rates. They have examined various
prescriptions with their work. Their prescription yields a better
agreement with the observations, but one can hardly consider it as
the final answer, especially for extreme differential rotation.

In general, what we can find from previous works on accretion
discs is the effects of various viscosity prescriptions on the
structure of discs and under conditions, their results are
compatible to some prescriptions. However, almost all previous
studies of $\beta$-discs dedicated to non-warped accretion discs.

In order to investigate the non-linear dynamics of warped
accretion discs and whether selecting of the model affects the
forms of the equations governing a warped viscous disc, we applied
Ogilvie's method (1999, hereafter OG) as considered in a $\alpha$
theory and check this new prescription for a thin viscous disc.
Our study includes numerical solutions of the Navier-Stokes
equations. First, the problem is reduced to a so-called singular
perturbation which is then solved by the method of matched
asymptotic expansions.

The structure of the paper is as the following. In the next
section, we explain general formulation. Analysis of the problem
is presented in section 3. Then we show that the equations can be
solved using the method of the separation of the variables and the
numerical solutions are discussed  in section 4. We compare the
warped $\alpha$ and $\beta$-discs in section 5  and then we conclude in section 6.\\
\section{General Formulation}
In order to construct a model for a warped accretion disc, we
start by writing hydrodynamic equations. Then, we use the
appropriate forms of these fundamental equations in warped
spherical polar coordinates $(r, \theta, \phi)$ (OG). Although
many physical agents such as magnetic fields and radiative
processes are playing significant roles in the dynamics of the
discs, we neglect all those complex phenomena in order to
understand the dynamics of an accretion disc with
$\beta$-prescription for viscosity via semi-analytical methods.
Also the self-gravity of the disc and interaction of the stellar
magnetic field with the disc are neglected. The fundamental
governing equations are the continuity,
\begin{eqnarray}
{\rm D}\rho=-\rho\nabla\cdot {\bf u},\label{drho}
\end{eqnarray}
the adiabatic condition,\begin{eqnarray}
{\rm D}p=-\Gamma p\nabla\cdot {\bf u},\label{dp}
\end{eqnarray}
and finally the equation of motion,\begin{equation}
\rho{\rm D} {\bf u}=-\nabla
p-\rho\nabla\Phi+\nabla\cdot\left[\mu\nabla {\bf u}+\mu(\nabla
{\bf u})^{\rm T}\right]+\nabla\left[(\mu_{\rm
b}-{\textstyle{{2}\over{3}}}\mu)\nabla\cdot {\bf u}\right]
,\label{du} \end{equation}
where $\bf u$, $\rho$, $p$ and $\Phi$ are the absolute velocity,
the density, the pressure and the external gravitational
potential, respectively. Also, $\Gamma$ is the adiabatic exponent
and the shear and the bulk viscosities are denoted by $\mu$ and
$\mu_{\rm b}$. The symbol $D$ denotes the Lagrangian time
derivative operator, \begin{eqnarray}
{\rm
D}=(\partial_t)_{r,\theta,\phi}+v_r\partial_r+{{v_\theta}\over{r}}
\partial_\theta+{{v_\phi}\over{r\sin\theta}}\partial_\phi.\label{d}
\end{eqnarray}
Note that the components $(u_{\rm r}, u_{\rm\theta}, u_{\rm\phi})$
are the absolute velocity components, i.e. $\bf u$ is the velocity
as measured in the inertial frame. But the additional motion with
respect to the warped coordinate system is described by the
relative velocity $\bf v$. The exact relationship between these
two velocities has been found by OG.

Warped spherical polar coordinates $(r, \theta, \phi)$ is defined
so that on each sphere $r=$ constant, one can define the usual
angular coordinates $(\theta, \phi)$, but with respect to an axis
that is tilted to a point in the direction of unit vector ${\bf
\ell}(r ,t)$. This tilt vector can be described using the Euler
angles $\beta_{E}(r,t)$ and $\gamma(r,t)$:\begin{eqnarray}
\ell=\sin\beta_{E}\cos\gamma\, {\bf
e}_x+\sin\beta_{E}\sin\gamma\,{\bf e}_y+\cos\beta_{E}\, {\bf
e}_z.\label{unit}\end{eqnarray}
For the viscosity coefficients, we are using the $\beta$-
prescription (Lynden-Bell and Pringle 1974) rather than the usual
$\alpha$- prescription which has been used by OG. Thus, the
viscosity coefficients  are assumed to be locally proportional to
the rotational velocity,\begin{eqnarray}
\mu&=&\beta r^2\Omega\rho,\label{mu}\\
\mu_{\rm b}&=&\beta_{\rm b}
r^2\Omega\rho,\label{mub}\end{eqnarray}
where the dimensionless coefficients $\beta$ and $\beta_{\rm b}$
can be considered as functions of the radius. We will show that
$\beta$- prescription for the viscosity coefficients lead to
significant changes in the equations describing a warped disc.
Also the fluid is assumed to be locally
polytropic,\begin{eqnarray}
p=k\rho^\Gamma.\label{p}\end{eqnarray}
where  $\Gamma(r)$ is a prescribed function of radius.

Now we can consider a thin disc in a spherically potential
$\Phi(r)$, in which the small parameter $\epsilon$ is a
characteristic value of the local angular semi-thickness of the
disc. Using this small parameter, it is possible to study the
structure of a warped disc by asymptotic expansion method. In thin
disc approximation, one can assume that the disc matter lies close
to $\theta=\pi/2$. To resolve the internal structure of the disc,
introduce the scaled dimensionless vertical coordinate
$\zeta$,\begin{equation}
\theta={{\pi}\over{2}}-\epsilon\zeta,\end{equation}
and the slow time coordinate,\begin{equation}
T=\epsilon^2t.\end{equation}
For the density and pressure, introduce the
scalings\begin{eqnarray}
\rho(r,\theta,\phi,t)&=&\epsilon^s\left[\rho_0(r,\phi,\zeta,T)+
\epsilon\rho_1(r,\phi,\zeta,T)+O(\epsilon^2)\right],\\
p(r,\theta,\phi,t)&=&\epsilon^{s+2}\left[p_0(r,\phi,\zeta,T)+\epsilon
p_1(r,\phi,\zeta,T)+O(\epsilon^2)\right],\end{eqnarray}
and for the relative velocities,\begin{eqnarray}
v_r(r,\theta,\phi,t)&=&\epsilon
v_{r1}(r,\phi,\zeta,T)+\epsilon^2v_{r2}(r,\phi,\zeta,T)+O(\epsilon^3),\\
v_\theta(r,\theta,\phi,t)&=&\epsilon
v_{\theta1}(r,\phi,\zeta,T)+\epsilon^2v_{\theta2}(r,\phi,\zeta,T)+
O(\epsilon^3),\\
v_\phi(r,\theta,\phi,t)&=&r\Omega(r)\sin\theta+\epsilon
v_{\phi1}(r,\phi,\zeta,T)+\epsilon^2
v_{\phi2}(r,\phi,\zeta,T)+O(\epsilon^3).\end{eqnarray}
Finally, for the viscosities, assume\begin{eqnarray}
\mu(r,\theta,\phi,t)&=&\epsilon^{s+2}\left[\mu_0(r,\phi,\zeta,T)+
\epsilon\mu_1(r,\phi,\zeta,T)+O(\epsilon^2)\right],\\
\mu_{\rm b}(r,\theta,\phi,t)&=&\epsilon^{s+2}\left[\mu_{{\rm
b}0}(r,\phi,\zeta,T)+\epsilon\mu_{{\rm
b}1}(r,\phi,\zeta,T)+O(\epsilon^2)\right].\end{eqnarray}
where $s$ is a parameter which should be positive if the
self-gravitation of the disc is to be negligible. The equations of
fluid dynamics were derived in warped spherical polar coordinates
by OG. He reduced them by means of above asymptotic expansions for
a thin disc and then divided them into two sets. Set A, which
determines the intermediate velocities, consists of five coupled
non-linear partial differential equations (PDEs) in two dimensions
$(\phi,\zeta)$ and seven dependent variables
$\{\rho_0,p_0,\mu_0,\mu_{b0},v_{r1},v_{\theta1},v_{\phi1}\}$. Also
Set B, which  determines the slow velocities, contains a set of
five linear PDEs for the higher-order quantities
$\{\rho_1,p_1,\mu_1,\mu_{b1},v_{r2},v_{\theta2},v_{\phi2}\}$,
including coefficients that depend upon the solutions of Set A and
their radial derivatives. In the present work, we adopted Set A
and B for our analysis (see OG, for details of deriving the expansions of Set A and Set B).\\
\section{Analysis}
It is very unlikely that the equations of Set A can be solved
analytically. Numerical approach is a convenient way. Fortunately,
we can transform the partial differential equations into a set of
ordinary differential equations (ODEs) using the method of
separation of variables. To achieve this, at first we introduce
the following forms for the variables:\begin{eqnarray}
h_0&=&r^2\Omega^2\left[f_1(\phi-\chi)-{\textstyle{{1}\over{2}}}
f_2(\phi-\chi)\zeta^2\right],\label{h0}\\
v_{r1}&=&r\Omega f_3(\phi-\chi)\zeta,\label{vr1}\\
v_{\theta1}&=&r\Omega\left[f_4(\phi-\chi)\zeta+g_4(\phi-\chi)\zeta^3\right],\label{vtheta1}\\
v_{\phi1}+rv_{r1}\gamma^\prime\cos\beta_E&=&r\Omega
\left[f_5(\phi-\chi)\zeta+g_5(\phi-\chi)\zeta^3\right].\label{vphi1}\end{eqnarray}
where $h_0={\Gamma\over{\Gamma-1}}{{p_0}\over{\rho_0}}$ is the
enthalpy. Using the relation (\ref{h0}), one can drive the upper
surface of the disc as \begin{eqnarray}
\zeta^2&=&2f_1(\phi-\chi)f_2^{-1}(\phi-\chi)\label{zeta2}\end{eqnarray}
Introducing by the dimensionless functions $f_1,\dots,f_5$ and
$g_4 ,g_5$, one can obtain the coupled sets of the non-linear ODEs
of first order for the given equations\begin{eqnarray}
f_1^\prime(\phi)&=&(\Gamma-1)f_4(\phi)f_1(\phi),\label{f1}\\
f_2^\prime(\phi)&=&(\Gamma+1)f_4(\phi)f_2(\phi)-6(\Gamma-1)g_4(\phi)f_1(\phi),\label{f2}\\
f_3^\prime(\phi)&=&f_4(\phi)f_3(\phi)+2f_5(\phi)+\left[f_2(\phi)-6(\beta_{\rm
b}+{\textstyle{{1}\over{3}}}\beta)g_4(\phi)\right]|\psi|\cos\phi\,\label{f3}\\
f_4^\prime(\phi)&=&-f_3^\prime(\phi)|\psi|\cos\phi+2f_3(\phi)|\psi|
\sin\phi+f_4(\phi)\left[f_4(\phi)+f_3(\phi)|\psi|\cos\phi\right]+1-f_2(\phi)\nonumber\\
&&+6(\beta_{\rm b}+{\textstyle{{1}\over{3}}}\beta)g_4(\phi)+6\beta
g_4(\phi)(1+|\psi|^2\cos^2\phi),\label{f4}\\
g_4^\prime(\phi)&=&g_4(\phi)f_3(\phi)|\psi|\cos\phi+4g_4(\phi)f_4(\phi),\label{g4}\\
f_5^\prime(\phi)&=&f_4(\phi)f_5(\phi)-{\textstyle{{1}\over{2}}}
\tilde\kappa^2f_3(\phi)+6\beta
g_5(\phi)(1+|\psi|^2\cos^2\phi),\label{f5}\\
g_5^\prime(\phi)&=&3f_4(\phi)g_5(\phi)+g_4(\phi)f_5(\phi),\label{g5}\end{eqnarray}
where these functions are subject to periodic boundary conditions
$f_n(2\pi)=f_n(0)$ and $g_n(2\pi)=g_n(0)$. The epicyclic frequency
$\kappa(r)$ is defined by \begin{equation}
\kappa^2=4\Omega^2+2r\Omega\Omega^\prime,\label{k2}
\end{equation}
and the dimensionless epicyclic frequency is
$\tilde\kappa=\kappa/\Omega$. Meanwhile the amplitude of the warp
is defined as\begin{eqnarray}
|\psi|&=&r|{{\partial\ell}\over{\partial
r}}|,\label{|psi|}\end{eqnarray}
hence, using the tilt vector, we have\begin{eqnarray}
\psi=|\psi|\,{\rm e}^{{\rm i}\chi}=r(\beta^\prime_E+\rm
i\gamma^\prime\sin\beta_E),\label{psi}\end{eqnarray}
that it is a dimensionless complex variable.\footnote{Throughout
this paper, a prime and a dot for $\Omega(r), \beta_E$ and
$\gamma$ imply radial and time derivatives, respectively.}
Moreover, by the definition of $f_6$ (see, Appendix B), one can
drive\begin{eqnarray}
f_6^\prime(\phi)=-2f_4(\phi)f_6(\phi)+9(\Gamma-1)f_1(\phi)f_2(\phi)^{-1}g_4(\phi)f_4(\phi).\label{f6}
\end{eqnarray}
Also in addition to these, we find the following
combinations\begin{eqnarray}
f_2(\phi)g_4(\phi)&=&0\qquad\hbox{if}\quad\Gamma\neq
1/3,\label{find1}\\
g_4(\phi)f_3(\phi)&=&-2g_5(\phi)\qquad\hbox{all}\quad\Gamma.\label{find2}\end{eqnarray}
All information about a warped disc with $\beta$- prescription can
be obtained by solving equations (\ref{f1})-(\ref{f6})
numerically. Before presenting the results of numerical
integration, we discuss about evolutionary equations of a warped
$\beta$- disc and its relations with solutions described by
functions $f_{n}$ and $g_n$. In particular, it is very important
to see whether the equations of Pringle (1992) can be derived from
the three dimensional fluid equations. Pringle (1992) developed an
approach for deriving the equations of a warped disc, without
reference to the detailed internal fluid equations. OG showed the
impossibility of deriving the angular momentum equation of Pringle
(1992) from the basic equations of fluid dynamics. In fact, one
should allow for a more general form of the torque between
neighboring rings. Thus, OG introduced three dimensionless
coefficients $Q_1$, $Q_2$ and $Q_3$, which depend on physical
quantities of the disc. These coefficients enable us to discuss
about relative importance of torques between neighboring rings in
a warped $\alpha$- disc. We are following similar approach for
describing a warped $\beta$- disc. However, as we will see, in a
warped $\beta$- disc there are various torques and so we should
introduce extra dimensionless coefficients.  Using the equations
that can be extracted from Set B in OG and substituting the
defined quantities $\{h_0,v_{r1},v_{\theta1},v_{\phi1}\}$, we can
propose the evolutionary equations for the warped $\beta$- disc as
(for details, see Appendix B),\begin{eqnarray}
\lefteqn{{{\partial}\over{\partial t}}\left(\Sigma r^2\Omega\ell
\right)+{{1}\over{r}}{{\partial}\over{\partial r}}\left(\Sigma\bar
v_rr^3\Omega\ell\right)=}&\nonumber\\
&&{{1}\over{r}}{{\partial}\over{\partial r}}\left(Q_1{\cal I}
r^2\Omega^2\ell \right)+{{1}\over{r}}{{\partial}\over{\partial
r}}\left(Q_2{\cal I} r^3\Omega^2{{\partial\ell}\over{\partial
r}}\right) +{{1}\over{r}}{{\partial}\over{\partial r}}(Q_3{\cal I}
r^3\Omega^2\ell\times{{\partial\ell}\over{\partial r}})\nonumber\\
&&+{{1}\over{r}}{{\partial}\over{\partial
r}}\left(Q^\prime_1\Sigma r^4\Omega^2\ell
\right)+{{1}\over{r}}{{\partial}\over{\partial
r}}\left(Q^\prime_2\Sigma r^5\Omega^2{{\partial
\ell}\over{\partial r}}
\right)+{{1}\over{r}}{{\partial}\over{\partial
r}}\left(Q^\prime_3\Sigma r^5\Omega^2\ell\times{{\partial
\ell}\over{\partial r}}\right),\label{anmo} \end{eqnarray}
for the angular momentum, and \begin{equation}
\Sigma\bar v_r{\partial\over{\partial
r}}(r^2\Omega)={{1}\over{r}}{{\partial}\over{\partial
r}}\left(Q_1{\cal I} r^2\Omega^2\right)-Q_2{\cal I}
r^2\Omega^2\left|{{\partial\ell}}\over{\partial
r}\right|^2+{{1}\over{r}}{{\partial}\over{\partial
r}}\left(Q^\prime_1\Sigma r^4\Omega^2\right)-Q^\prime_2\Sigma
r^4\Omega^2\left|{{\partial\ell}}\over{\partial
r}\right|^2,\label{anmol}\end{equation}
for the component of angular momentum parallel to $\ell$, and
\begin{eqnarray}
\lefteqn{\Sigma r^2\Omega\left({{\partial\ell}\over{\partial
t}}+\bar v_r{{\partial\ell}\over{\partial
r}}\right)=}&\nonumber\\
&&Q_1{\cal I} r\Omega^2{{\partial\ell}\over{\partial
r}}+{{1}\over{r}}{{\partial}\over{\partial r}}\left(Q_2{\cal I}
r^3\Omega^2{{\partial\ell}\over{\partial r}}\right)+Q_2{\cal I}
r^2\Omega^2\left|{{\partial\ell}\over{\partial
r}}\right|^2{\ell}+{{1}\over{r}}{{\partial}\over{\partial
r}}\left(Q_3{\cal I}
r^3\Omega^2{\ell}\times{{\partial\ell}\over{\partial
r}}\right)\nonumber\\
&& +Q^\prime_1\Sigma r^3\Omega^2{{\partial\ell}\over{\partial
r}}+{{1}\over{r}}{{\partial}\over{\partial
r}}\left(Q^\prime_2\Sigma r^5\Omega^2{{\partial\ell}\over{\partial
r}}\right)+Q^\prime_2\Sigma
r^4\Omega^2\left|{{\partial\ell}\over{\partial
r}}\right|^2{\ell}+{{1}\over{r}}{{\partial}\over{\partial
r}}\left(Q^\prime_3
r^5\Omega^2{\ell}\times{{\partial\ell}\over{\partial
r}}\right).\label{tivec}\nonumber\\
\end{eqnarray}
for the tilt vector.\\
Thus, we can obtain coefficients $Q_n$ and $Q^\prime_n$
$\{i.e.,n=1,4\}$ as (Appendix B), \begin{eqnarray}
Q_1&=&\big\langle f_6(-f_3f_5+3\beta g_5|\psi|\cos\phi)\big\rangle,\label{q1}\\
Q^\prime_1&=&\big\langle
-{\textstyle{{1}\over{2}}}(4-\tilde\kappa^2)\beta
+\beta f_5|\psi|\cos\phi\big\rangle,\label{q'1}\\
Q_4&=&{1\over{|\psi|}}\big\langle{\rm e}^{{\rm
i}\phi}f_6\left[f_3-{\rm i}f_3(f_4+f_3|\psi|\cos\phi)+3{\rm
i}\beta g_4|\psi|\cos\phi\right]\big\rangle,\label{q4}\\
Q^\prime_4&=&{1\over{|\psi|}}\big\langle{\rm e}^{{\rm
i}\phi}\left[{\rm i}\beta
(f_4+f_3|\psi|\cos\phi)|\psi|\cos\phi-{\rm i}\beta f_3-{\rm
i}\beta|\psi|\sin\phi\right]\big\rangle,\label{q'4} \end{eqnarray}
where $Q_4$ and $Q^\prime_4$  defined as  $Q_2+{\rm i}Q_3$ and
$Q^\prime_2+{\rm i}Q^\prime_3$.\\

The dimensionless coefficients $Q_n$ and $Q^\prime_n$ can be
evaluated from the solutions of Set A (see OG). We find that these
coefficients depend on selecting the model, the amplitude of the
warp, the rotation law and the shear viscosity. For a warped
$\beta$-disc as we consider here, we find three kinds of the
internal viscous torques, (see equation (\ref{anmo})):\beqa
G(r,t)&=&(Q_1{\cal I}+Q^\prime_1\Sigma
r^2)r^2\Omega^2\ell+(Q_2{\cal I}+Q^\prime_2\Sigma
r^2)r^3\Omega^2\frac{\partial\ell}{\partial
r}\nonumber\\
&&+(Q_3{\cal I}+Q^\prime_3\Sigma
r^2)r^3\Omega^2\ell\times\frac{\partial\ell}{\partial
r}.\nonumber\\
\eeqa
From a mathematical point of view, the equation (\ref{anmo})
constitutes the prototype for a parabolic partial differential
equation and can be thought as an advection-diffusion-dispersion
equation in the non-linear regime. So for a flat disc
$(\frac{\partial\ell}{\partial r}=0)$, the evolution equation
reduces to a standard disc diffusion equation.

According to the above equation for  G(r,t), the first term on the
right hand side gives a contribution to G which is in the local
direction $\ell$. So, the effective advection coefficients, $Q_1$
and $Q^\prime_1$, represent viscous torques on each ring in the
disc due to differential rotation within the disc plane. Thus, the
rings tend to rotate in direction $\ell$.

Considering neighboring rings in directions $\ell$ and
$\ell+\Delta\ell$, the second term on the right hand side gives a
contribution to G which is in the $\frac{\partial\ell}{\partial
r}$ direction. Therefore, the effective diffusion coefficients,
$Q_2$ and $Q^\prime_2$, represent viscous torques due to the
interaction between neighboring rings in the disc, in order to
flatten the disc.

The last term on the right-hand side is a dispersion one. So the
effective dispersion coefficients $Q_3$ and $Q^\prime_3$
demonstrate torques which are perpendicular to $\ell$ and
$\frac{\partial\ell}{\partial r}$, respectively. In this case,
each ring in the disc experiences torques tending to make the ring
precess if it is not aligned with its neighbors. Then we would
expect to generate wave motions in the disc so that the warp
propagates as a dispersive wave.

\section{Numerical solutions}
One can expand the dimensionless functions $f_n$
$\{n=1,\dots,6\}$, $g_4$ and $g_5$ as power series of the
amplitude of the warp to determine the dynamics to any desired
order. In Appendix A, truncated Taylor series are presented
including the coefficients $Q_n$ and $Q^\prime_n$. These series
enable us to start numerical integration. From the expansions of
the dimensionless functions $f_n$ and $g_n$ and their relations to
coefficients $Q_n$ and $Q^\prime_n$, we found truncated Taylor
series only for coefficients $Q^\prime_1,Q^\prime_2$ and $Q_3$.
Meanwhile, as we discuss in Appendix A, our solutions are
restricted to the nearly Keplerian-discs. So we consider in our
calculations for $\tilde\kappa^2=0.99$ for all values of $\beta$.

In order to be able to compare  our results with the warped
$\alpha$- disc, we adopt coefficients $Q_1$ and $Q_4=Q_2+{\rm
i}Q_3$ from OG. To distinct between these two coefficients, we
label them by superscript $\alpha$. Now the set of ODEs in the
previous section can be solved numerically and we may obtain the
coefficients $Q^\prime_n$ and $Q_n$. In addition, we should note
that solutions must satisfy periodic conditions for functions
$f_n$ and $g_n$ as well as $\big\langle f_6\big\rangle=1$ (see,
Appendix B). Thus, the derived solutions
make an interesting dynamical description of a warped $\beta$- disc. \\

\subsection{The warped $\beta$-disc}
First, we investigate a non-Keplerian disc without viscosity. We
consider $\Gamma=5/3$ and $\beta=\beta_b=0$. The dimensionless
functions $f_1,f_2,f_5,f_6$ and $g_4$ are even, but $f_3,f_4$ and
$g_5$ are odd. So the only non vanishing coefficient is $Q_3$.
So,\begin{eqnarray}
Q_3&=&{1\over{|\psi|}}\big\langle
f_6[f_3\sin\phi-f_3\cos\phi(f_4+f_3|\psi|\cos\phi)+ 3\beta
g_4|\psi|\cos^2\phi]\big\rangle\mid_{\beta=0}\nonumber,\\
&=&{1\over{|\psi|}}\big\langle
f_6[f_3\sin\phi-f_3\cos\phi(f_4+f_3|\psi|\cos\phi)]\big\rangle.\label{q3}\end{eqnarray}

A contour plot of coefficient $Q_3$, is shown in Figure 1.
\begin{center}
  \includegraphics[width=3.5583in]{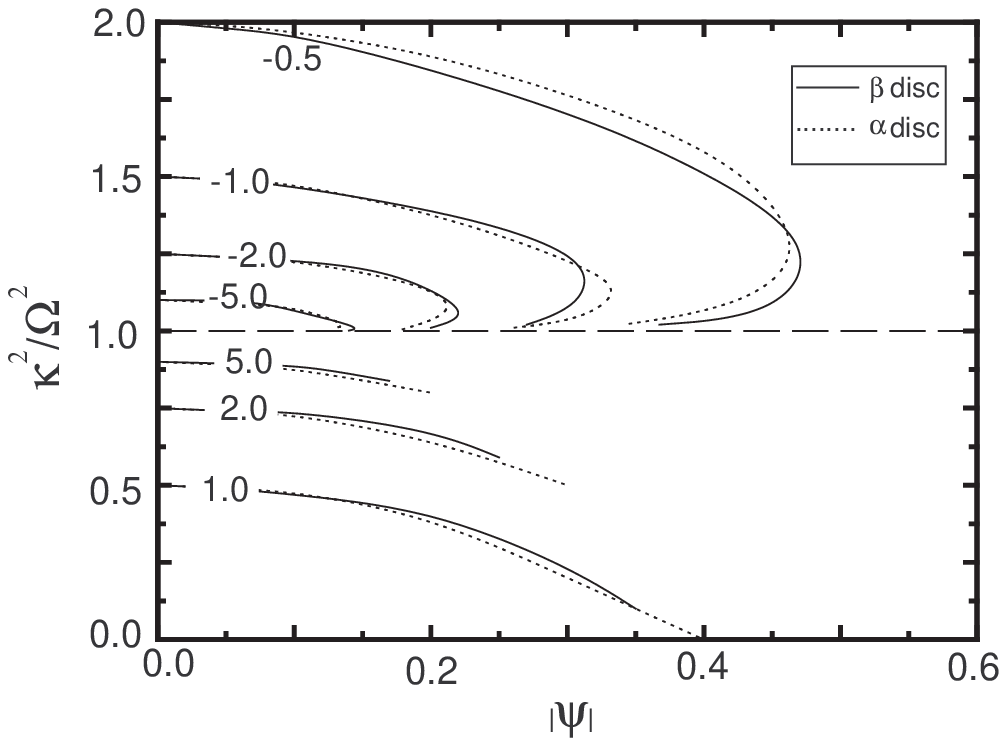}\\
\end{center}
\noindent \emph {Figure 1.The square of  dimensionless epicyclic
frequency versus the amplitude of the warp for Contour plots of
the coefficients $Q_3$(solid line) and $Q^\alpha_3$ (dash line)
for an inviscid disc with $\Gamma=5/3$.The condition
$\tilde\kappa^2=1$ separates solutions into two parts. For
$\tilde\kappa^2<1$, we have $Q_3,Q^\alpha_3>0$; and for
$\tilde\kappa^2>1$, we see $Q_3,Q^\alpha_3<0$.}

For the case of nearly Keplerian disc with viscosity, we consider
$\tilde\kappa^2=0.99$, $\Gamma=5/3$ and $\beta_b=0$. Figure 2,
shows $\beta$ vs. $|\psi|$ and contours of all coefficients. Note
that all the plots in Figure 2; parts (a), (c) and (e) show
coefficients $Q^\prime_n$ and parts (b), (d) and (f) show
coefficients $Q_n$ . The solutions can be calculated for large
values of $|\psi|$. For evaluating the coefficients $Q_1$ and
$Q^\prime_1$, we used equations (\ref{q1}) and (\ref{q'1}).
Figures 2a and 2b show contour of plots of the coefficients
$Q^\prime_1$ and $Q_1$ respectively. For reasonably small values
of $|\psi|$ there is a good agreement with the truncated Taylor
series for $Q^\prime_1$.

Figures 2c and 2d show contour plots of the coefficients
$Q^\prime_2$ and $Q_2$ respectively. We used equations (\ref{q4})
and (\ref{q'4}) for evaluating the coefficients $Q_2$ and
$Q^\prime_2$, respectively. For reasonably small values of
$|\psi|$ there is a good agreement with the truncated Taylor
series for $Q^\prime_2$.

Figures 2e and 2f show contours  of the coefficients $Q^\prime_3$
and $Q_3$ respectively. We used equations (\ref{q4}) and
(\ref{q'4}) for evaluating the coefficients $Q_3$ and
$Q^\prime_3$, respectively. Also $Q^\prime_3$ is much smaller in
magnitude than $Q_3$. For reasonably small values of $|\psi|$
there is a good agreement with the truncated Taylor series for $Q_3$.\\
\begin{center}
  \includegraphics[width=5.5583in]{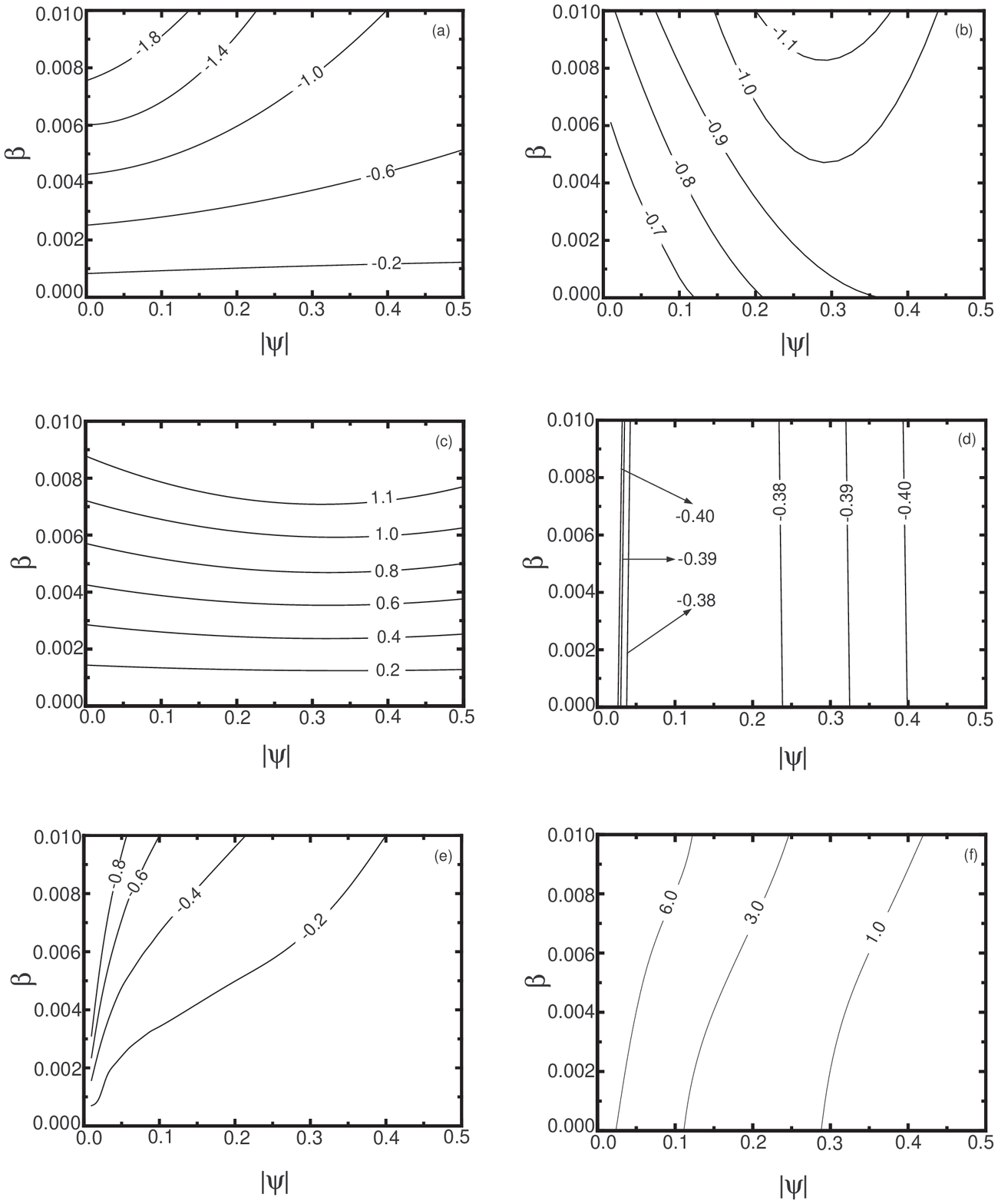}\\
\end{center}
\noindent \emph {Figure2. The dimensionless viscosity parameter
versus the amplitude of the warp for Contour plots of the
coefficients; (a)$Q^\prime_1$, (b)$Q_1$, (c)$Q^\prime_2$,
(d)$Q_2$, (e)$Q^\prime_3$  and (f)$Q_3$ for a viscous, nearly
Keplerian disc ($\tilde\kappa^2=0.99$)with $\Gamma=5/3$ and
$\beta_{\rm b}=0$.}
\subsection{The warped $\alpha$-disc}
First, as an illustrative case, we investigate a non-Keplerian
disc without viscosity. We consider $\Gamma=5/3$ and
$\alpha=\alpha_b=0$. According to obtained functions
$f_1,f_2,f_5,f_6$ by OG, the only non vanishing coefficient is
$Q^\alpha_3$. So,\begin{eqnarray}
Q_3^\alpha&=&{1\over{|\psi|}}\big\langle
f_6[f_3\sin\phi-f_3\cos\phi(f_4+f_3|\psi|\cos\phi)]\big\rangle.\label{qinvalpha3}
\end{eqnarray}
If we plot $\kappa^2/\Omega^2$ versus $|\psi|$, we get the
contours indicated in Figure 1 for $Q^\alpha_3$. We selected a
given interval for $|\psi|$.

For the case of nearly Keplerian disc with viscosity, we consider
$\tilde\kappa^2=0.99$, $\Gamma=5/3$ and $\alpha_b=0$. Figure 3
shows $\alpha$ vs. $|\psi|$ and contours of all coefficients. Note
that all the plots in Figure 3; parts (a), (b) and (c) show
coefficients $Q^\alpha_n$. We adopt this coefficients from OG,
\begin{eqnarray}
Q^\alpha_1&=&\big\langle f_6[{-1\over{2}}(4-\tilde\kappa^2)\alpha
f_2-f_3f_5+\alpha f_2f_5|\psi|\cos\phi]\big\rangle.\label{qalpha1}
\end{eqnarray}
\begin{eqnarray}
Q^\alpha_2&=&{1\over{|\psi|^2}}\big\langle
f_6[(f_4+f_3|\psi|\cos\phi)(1+f_3|\psi|\sin\phi)+\alpha
f_2f_3|\psi|\sin\phi\nonumber\\
&&-\alpha
f_2(f_4+f_3|\psi|\cos\phi)|\psi|^2\cos\phi\sin\phi+\alpha
f_2|\psi|^2\sin^2\phi]\big\rangle.\label{qalpha2}
\end{eqnarray}
\begin{eqnarray}
Q^\alpha_3&=&{1\over{|\psi|}}\big\langle
f_6[f_3\sin\phi-f_3\cos\phi(f_4+f_3|\psi|\cos\phi)+\alpha
f_2(f_4+f_3|\psi|\cos\phi)|\psi|\cos^2\phi\nonumber\\
&&-\alpha f_2f_3\cos\phi-\alpha
f_2|\psi|\sin\phi\cos\phi]\big\rangle.\label{qvalpha3}
\end{eqnarray}
\section{Comparing the warped $\alpha$ and $\beta$-discs}
\subsection{Inviscid, non-Keplerian disc}
Figure 1 shows a good agreement between two contour plots of $Q_3$
and $Q^\alpha_3$. Also they show good agreements with truncated
Taylor series for small values of $|\psi|$. As Figure 1
demonstrates the condition $\tilde\kappa^2=1$ separates solutions
into two parts. For $\tilde\kappa^2<1$, we have
$Q_3,Q^\alpha_3>0$; and for $\tilde\kappa^2>1$, we see
$Q_3,Q^\alpha_3<0$. In the case $\tilde\kappa^2<1$ and for small
values of $|\psi|$, the physical solutions of $\beta$ and $\alpha$
models exist only for $|\psi|<0.35$ and $|\psi|<0.4$ respectively.
The solutions terminates when $g_4=0,f_2\ll1$ for $\beta$ model
because we couldn't obtain the numerical solution. Whereas the
solutions terminate for $\alpha$ model when $f_2=0$. Hence, it
seems that the stability condition establishes for such discs for
all values $\tilde\kappa^2>0$ in small values of $|\psi|$. In the
other word, under the condition $\tilde\kappa^2>0$, since the
epicyclic frequency has the relation with particle orbits, the
displacements oscillate about a fixed mean position and the
circular orbit is stable to small perturbations.\\
\begin{center}
  \includegraphics[width=2.5583in]{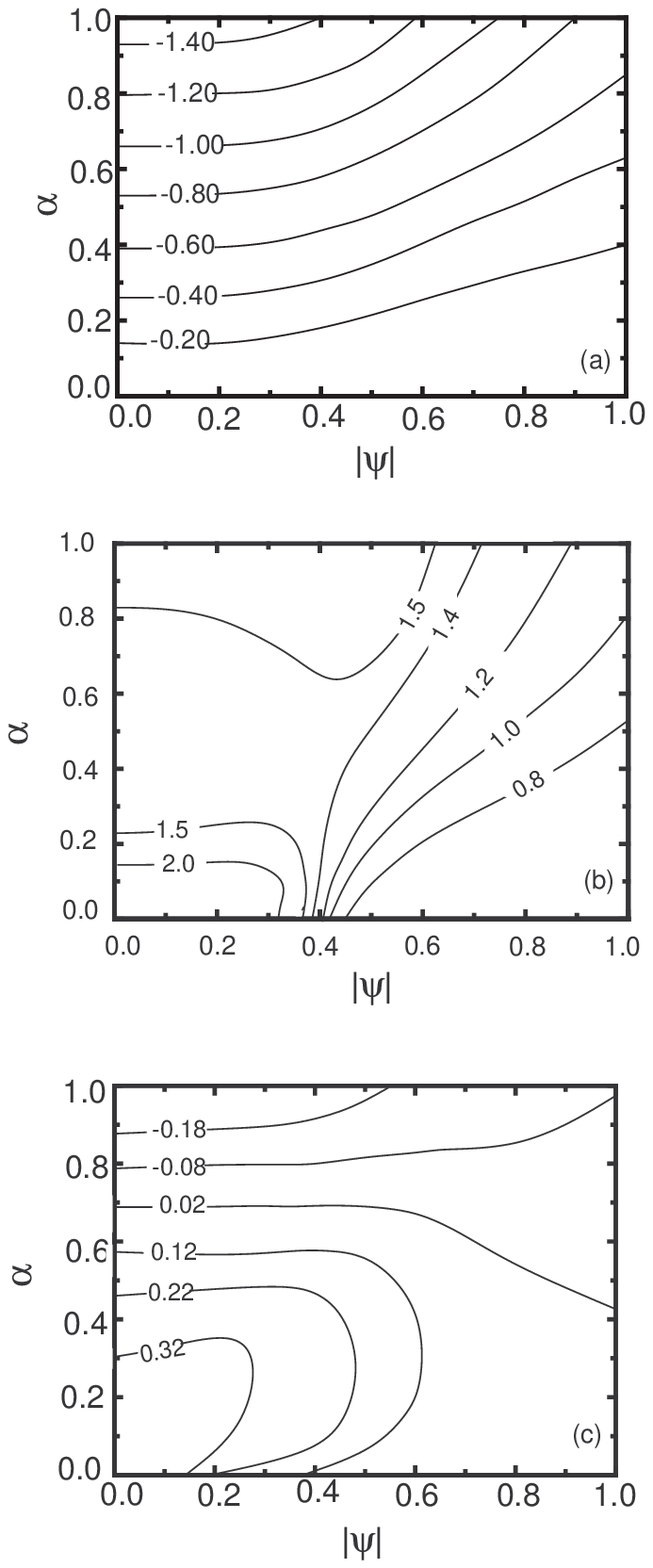}\\
\end{center}
\noindent \emph {Figure 3.The dimensionless viscosity parameter
versus the amplitude of the warp for Contour plots of the
coefficients; (a)$Q^\alpha_1$, (b)$Q^\alpha_2$ and (c)$Q^\alpha_3$
for a viscous, nearly Keplerian disc ($\tilde\kappa^2=0.99$)with
$\Gamma=5/3$ and $\alpha_{\rm b}=0$.}
\subsection{Viscous, nearly Keplerian disc}
Now, we  compare $\alpha$ and $\beta$ models for a viscous, nearly
Keplerian disc. If we compare the equation (\ref{anmo}) with
Pringle's equation (\ref{L1}), the coefficients $Q^\prime_1$ and
$Q_1$ represent viscous torques according to horizontal shear.
Figures 2a and 2b show similar behaviors, the magnitudes of
$Q^\prime_1$ and $Q_1$ increase with increasing $|\psi|$ and this
is similar to $Q^\alpha_1$ (Figure 3a). Nevertheless, we see
different treatments between two models when torques due to
$Q^\prime_1$ and $Q_1$ are considered. As mentioned before, they
indicate viscous torques parallel to $\ell$ and $Q^\prime_1$ ,
$Q_1$ identified as the advection coefficients. Then comparing to
$\alpha$ model, the advection of the warp occurs more
significantly. On the other hand, the viscous stability condition
implies that
 $Q_1$, $Q^\prime_1<0$, so $\beta$-discs comparing to
$\alpha$-discs tend to become stable more quickly. This is in
agreement to Hure et al. (2001) based on viscous stability for
$\beta$-discs which is one of the important property of them.
Thus, they do not tend to fragment in contrary to $\alpha$-disc.

Also,  the coefficients $Q^\prime_2$ and $Q_2$ represent viscous
torques due to the vertical shear. Figures 2c and 3b show
different behaviors. We see that  with increasing $|\psi|$,
$Q^\prime_2$ increases whereas $Q^\alpha_2$ decreases. So, the
results due to the torques of $Q^\prime_2$ and $Q_2$ don't tend to
OG's results. As if, this model affects the vertical structure of
$\alpha$-disc so that viscous torques play more role to flatten
$\beta$-disc. Also comparing to the OG's analysis, the
coefficients of $Q^\prime_3$ and $Q_3$ represent torques which
lead to the dispersive wave-like propagation of the warp. Figures
2e, 2f and 3c show that $Q^\prime_3$, $Q_3$ and $Q^\alpha_3$
respectively, they treat in a similar manner so that with
increasing $|\psi|$, they decrease. Although their typical
treatments are similar,   $Q_3$ is typically much larger than
$Q^\prime_3$ and $Q^\alpha_3$ coefficients especially in small
$|\psi|$. So we expect to generate wave motion more in small
$|\psi|$ for $\beta$-disc so that it loses its importance for
large $|\psi|$.
\section{Summary and Discussion}
We have presented an analysis of the non-linear dynamics of a
warped accretion disc using $\beta$-prescription. Using basic
equations of the fluid dynamics in a warped spherical polar
coordinates, we obtained the general equations that describe a
warped disc. We have also employed the method of matched
asymptotic expansions for thin discs to derive a set of coupled
PDEs which govern the dynamics of the system. Then we solved the
equations by the method of the separation of variables in order to
extracting the equations governing the warp in their simplest
forms. The non-linear dynamics of a warped $\beta$- disc under a
differentially rotation field in the presence of a spherically
symmetric external potential presents a more complicated behavior
comparing to the case in which a warped $\alpha$- disc is
considered (OG). Moreover, it is more complicated than the case of
the linear theory of Papaloizou \& Pringle (1983) to allow an
arbitrary rotation law. In our study, there are six coefficients.
They have been determined numerically and analytically, by solving
a set of ODEs. Using truncated Taylor series in the amplitude of
the warp, the coefficients  $Q^\prime_1$, $Q^\prime_2$ and $Q_3$
are calculated analytically. Figures 1-3, show their contour plots
for two cases inviscid-Keplerian and viscous-nearly Keplerian disc
when $\Gamma=5/3$ and $\beta_b=0$. We found that these
coefficients depend on selecting of the model and the shear
viscosity. Considering the equations (\ref{q1})...(\ref{q'4}), one
can easily note that  the coefficients also depend on $\Gamma$,
$\beta_b$ and the value of $|\psi|$.
Our results show that the equations governing a warp viscous disc
depend on the parameters of the model. The results can be
compared with the studies of Pringle (1992) and OG as follows:\\
\\
1)A comparison with the notation of Pringle (1992), there are four
different viscous torques due to the interaction between
neighboring rings in the disc (see, Appendix B). Therefore, these
torques may be explained by four coefficients. Comparing to OG, we
will find two extra coefficients $Q^\prime_1$ and $Q^\prime_2$ for
the viscous torques. To understand the meaning of these
coefficients, we may deduce following equations by comparing
equations (\ref{L1}) and (\ref{L4}),\begin{eqnarray}
Q_1{\cal I}\Omega+Q^\prime_1\Sigma
r^2\Omega&\longleftrightarrow&\nu_1\Sigma{d\ln\Omega\over{d\ln
r}},\label{compequ1}\\
Q_2{\cal I}\Omega+Q^\prime_2\Sigma
r^2\Omega&\longleftrightarrow&{1\over{2}}\nu_2\Sigma.\label{compequ2}\end{eqnarray}
We see that the numerical evaluation of the coefficients $Q_1,
Q^\prime_1$ and $Q_2, Q^\prime_2$ represents the qualitative
behavior of $\nu_1$ and $\nu_2$. We may conclude that $\nu_1$ and
$\nu_2$ depend on the value of $\beta$.\\
\\
2)The evolutionary equations for the warped disc show that this
scheme is a generalization of the form suggested by OG. A warped
accretion disc can be studied in detail by the coefficients
$Q^\prime_3$ and $Q_3$ in $\beta$ model and only the coefficient
$Q^\alpha_3$ in $\alpha$ model (OG). These coefficients
demonstrate torques tending to make the ring precess if it is
misaligned with its neighbors. Comparing Figures 2e and 3c show
that the direction of the torques is negative for both
coefficients $Q^\prime_3$, $Q^\alpha_3$. These torques lead to the
dispersive wave-like propagation of the warp.\\
\\
3)Our results show that different viscosity prescriptions and
magnitudes ($\alpha$ and $\beta$ prescriptions) affect dynamics of
a warped accretion disc. Therefore, it can be important in
determining the viscosity law even for a warped disc.
\\
Some further work still remains to study the dynamics of a viscous
warped accretion disc using $\beta$-prescription:\\
\\
a)The dynamics of the warped accretion disc may be studied by
considering the magnetic effects. Then we have to modify the
equations by including magnetic field of the disc. Thus, one
should introduce a suitable model for the geometry of the magnetic
field components. Also, one can extend our analysis to the case
that self-gravity of the warped discs is important.\\
\\
b)In this paper, we neglected the thermal and the radiative
effects, however, these physical processes are playing important
roles in the dynamics of the discs. Therefore, the behavior of the
dynamics of
these discs needs to be studied further.\\
\\
c)As we discussed, the expansions fail only when
$\tilde\kappa^2=1$ for all values of $\beta$. This is a resonant
case, which can't be described using this method. Therefore, it is
interesting to study the non-linear dynamics of the resonant
case for a viscous Keplerian (or nearly Keplerian) disc.\\
\\

\newpage
\appendix
\section{Truncated non-linear equations}
In this Appendix we find truncated Taylor series of the
dimensionless functions $f_{n} \{n=1,\dots,6\}$, $g_4$, $g_5$ and
the coefficients $Q_1,Q^\prime_1, Q_4$ and $Q^\prime_4$ in terms
of powers of $|\psi|$. Thus,\begin{eqnarray}
f_1(\phi)&=&f_{10}+|\psi|^2f_{12}(\phi)+O(|\psi|^4),\\
f_2(\phi)&=&f_{20}+|\psi|^2f_{22}(\phi)+O(|\psi|^4),\\
f_3(\phi)&=&|\psi|f_{31}(\phi)+|\psi|^2f_{32}(\phi)+|\psi|^3f_{33}(\phi)+O(|\psi|^4),\\
f_4(\phi)&=&|\psi|^2f_{42}(\phi)+O(|\psi|^4),\\
g_4(\phi)&=&|\psi|g_{41}(\phi)+|\psi|^3g_{43}(\phi)+O(|\psi|^5),\\
f_5(\phi)&=&|\psi|f_{51}(\phi)+|\psi|^2f_{52}(\phi)+|\psi|^3f_{53}(\phi)+O(|\psi|^4),\\
g_5(\phi)&=&|\psi|^2g_{52}(\phi)+|\psi|^3g_{53}(\phi)+O(|\psi|^4),\\
f_6(\phi)&=&f_{60}+|\psi|^2f_{62}(\phi)+O(|\psi|^4).\end{eqnarray}
Note  all terms that are scaled with $|\psi|^0$ indicate
an unwarped disc.\\

\subsection{Zeroth-order solution}
For an unwarped disc, equation (\ref{f4}) at $O(|\psi|^0)$
gives\begin{eqnarray}
f_{20}&=&1.\end{eqnarray}
Also equation (\ref{f6}) at $O(|\psi|^0)$ yields
\begin{eqnarray}
f_{60}^\prime(\phi)&=&1,\end{eqnarray}
and since $\langle f_6\rangle=1$ by definition (see, Appendix B),
then \begin{eqnarray}
f_{60}&=&1.\end{eqnarray}
\subsection{First-order solution}
The vertical velocity at first order is obtained using equation
(\ref{g4}) at $O(|\psi|)$,\begin{eqnarray}
g_{41}^\prime(\phi)&=&0,\eeqa
Hence\begin{eqnarray}
g_{41}&=&\tilde C_{\theta1}.\eeqa
Also the horizontal velocities at first order are obtained by
equations (\ref{f3}) and (\ref{f5}) at
$O(|\psi|)$,\begin{eqnarray}
f_{31}^\prime(\phi)-2f_{51}(\phi)&=&\cos\phi,\\
f_{51}^\prime(\phi)+{\textstyle{{1}\over{2}}}\tilde\kappa^2f_{31}(\phi)&=&0.\eeqa
Therefore, \begin{eqnarray}
f_{31}(\phi)&=&C_{r1}\cos\phi+S_{r1}\sin\phi,\\
f_{51}(\phi)&=&C_{\phi1}\cos\phi+S_{\phi1}\sin\phi.\eeqa
We can express the solutions using a complex notation as $Z=C+{\rm
i}S$,\begin{eqnarray}
\left[\matrix{-{\rm i}&-2\cr
{\textstyle{{1}\over{2}}}\tilde\kappa^2&-{\rm
i}\cr}\right]\left[\matrix{Z_{r1}\cr
Z_{\phi1}\cr}\right]=\left[\matrix{1\cr
0\cr}\right].\label{zrzphi1 }\eeqa
The determinant of this matrix is $-(1-\tilde\kappa^2)$, and
therefore, there isn't any solution for the Keplerian-disc
($\tilde\kappa^2=1$). Hence,we have\begin{eqnarray}
Z_{r1}&=&{{\rm i}\over{1-\tilde\kappa^2}},\\
Z_{\phi1}&=&{\tilde\kappa^2\over{2(1-\tilde\kappa^2)}}.\eeqa\\
\subsection{Second-order solution}
The horizontal velocities at second order are obtained by
equations (\ref{f3}), (\ref{f5}) and (\ref{g5}) at
$O(|\psi|^2)$\begin{eqnarray}
f_{32}^\prime(\phi)-2f_{52}(\phi)&=&-6(\beta_{\rm
b}+{\textstyle{{1}\over{3}}}\beta)g_{41}\cos\phi,\label{f32}\\
f_{52}^\prime(\phi)+{\textstyle{{1}\over{2}}}\tilde\kappa^2f_{32}(\phi)&=&6\beta
g_{52}(\phi),\label{f52}\\
g_{52}^\prime(\phi)&=&g_{41}f_{51}(\phi).\label{g52}\eeqa
Equation (\ref{g52}) yields a solution as\begin{eqnarray}
g_{52}(\phi)&=&\tilde C_{\theta1}Z_{\phi1}\sin\phi\eeqa
while equations (\ref{f32}) and (\ref{f52}) have the solutions as
follows\begin{eqnarray}
f_{32}(\phi)&=&C_{r2}\cos\phi+S_{r2}\sin\phi,\\
f_{52}(\phi)&=&C_{\phi2}\cos\phi+S_{\phi2}\sin\phi,\eeqa
with\begin{eqnarray}
\left[\matrix{-{\rm i}&-2\cr
{\textstyle{{1}\over{2}}}\tilde\kappa^2&-{\rm
i}\cr}\right]\left[\matrix{Z_{r1}\cr
Z_{\phi1}\cr}\right]=\left[\matrix{-6(\beta_{\rm
b}+{\textstyle{{1}\over{3}}}\beta)\tilde C_{\theta1}\cr
6{\rm i}\beta Z_{\phi1}\tilde C_{\theta1}\cr}\right].\label{zrzphi2}\eeqa\\
so the solution is\begin{eqnarray}
Z_{r2}&=&6{\rm i}\tilde C_{\theta1}{{(\beta_{\rm
b}+{\textstyle{{1}\over{3}}}\beta)+2\beta
Z_{\phi1}}\over{\tilde\kappa^2-1}},\eeqa\begin{eqnarray}
Z_{\phi2}&=&3\tilde C_{\theta1}{{\tilde\kappa^2(\beta_{\rm
b}+{\textstyle{{1}\over{3}}}\beta)+2\beta
Z_{\phi1}}\over{\tilde\kappa^2-1}}.\eeqa
Also the enthalpy and vertical velocity at the second order are
obtained by equation (\ref{f1}) at $O(|\psi|^2)$,
\begin{eqnarray}
f_{12}^\prime(\phi)&=&(\Gamma-1)f_{42}(\phi)f_{10},\eeqa
equation (\ref{f2}) at $O(|\psi|^2)$,\begin{eqnarray}
f_{22}^\prime(\phi)&=&(\Gamma+1)f_{42}(\phi)-6(\Gamma-1)f_{10}g_{41},\eeqa
and equation (\ref{f4}) at $O(|\psi|^2)$\begin{eqnarray}
f_{42}^\prime(\phi)&=&-f_{31}^\prime(\phi)
\cos\phi+2f_{31}(\phi)\sin\phi-f_{22}(\phi).\eeqa
Combining the last two equations, gives\begin{eqnarray}
f_{42}^{\prime\prime}(\phi)+(\Gamma+1)f_{42}(\phi)&=&3S_{r1}\sin2\phi.\eeqa
It has a solution as the following form\begin{eqnarray}
f_{42}(\phi)&=&C_{\theta2}\cos2\phi+S_{\theta2}\sin2\phi,\eeqa
with\begin{equation}
\left[\matrix{-(3-\Gamma)&0\cr 0&-(3-\Gamma)\cr}\right]\left[
\matrix{C_{\theta2}\cr S_{\theta2}\cr}\right]=\left[\matrix{0\cr
3S_{r1}\cr}\right].\eeq
In a complex notation, the solution is\begin{eqnarray}
Z_{\theta2}&=&{3{\rm i}\over{\Gamma-3}}S_{r1}.\eeqa
It then follows that\begin{eqnarray}
f_{12}(\phi)&=&-{\textstyle{{1}\over{2}}}(\Gamma-1)f_{10}S_{\theta2}
\cos2\phi,\\
f_{22}(\phi)&=&-{\textstyle{{1}\over{2}}}(\Gamma+1)S_{\theta2}\cos2\phi+{{\Gamma}\over{\Gamma-3}}S_{r1},\\
 f_{62}(\phi)&=&S_{\theta2}\cos2\phi.\eeqa\\
\subsection{Third-order solution}
The horizontal velocities at third order are obtained by equations
(\ref{f3}), (\ref{f5}) and (\ref{g5}) at
$O(|\psi|^3)$,\begin{eqnarray}
f_{33}^\prime(\phi)-2f_{53}(\phi)&=&f_{42}(\phi)f_{31}(\phi)+f_{22}(\phi)\cos\phi,\label{f33}\\
f_{53}^\prime(\phi)+{\textstyle{{1}\over{2}}}\tilde\kappa^2f_{33}(\phi)&=&
f_{42}(\phi)f_{51}(\phi)+6\beta g_{53}(\phi),\label{f53}\\
g_{53}^\prime(\phi)&=&g_{41}f_{52}(\phi).\label{g53}\eeqa
Equation (\ref{g53}) yields a solution as\beqa
g_{53}(\phi)&=&\tilde C_{\theta1}Z_{\phi2}\sin\phi\eeqa
while equations (\ref{f33}) and (\ref{f53}) have the solutions as
follows\beqa
f_{33}(\phi)&=&C_{r3}\cos\phi+S_{r3}\sin\phi+\{\hbox{$m=3$ terms}\},\\
f_{53}(\phi)&=&C_{\phi3}\cos\phi+S_{\phi3}\sin\phi+\{\hbox{$m=3$terms}\}.\eeqa
hence,one can obtain \beqa
Z_{r3}&=&{{\rm
i}\over{\tilde\kappa^2-1}}\left[-{{\Gamma}\over{\Gamma-3}}S_{r1}
+{{\Gamma+1}\over{4}}S_{\theta2}-{{1}\over{2}}S_{\theta2}S_{r1}+
S_{\theta2}Z_{\phi1}+12\beta\tilde C_{\theta1}Z_{\phi2}\right],\\
Z_{\phi3}&=&{{1}\over{\tilde\kappa^2-1}}\left[{{1}\over{2}}S_{\theta2}C_{\phi1}+
6\beta\tilde
C_{\theta1}Z_{\phi2}-{{\tilde\kappa^2}\over{2}}\left({{\Gamma}\over{\Gamma-3}}S_{r1}
-{{\Gamma+1}\over{4}}S_{\theta2}+{{1}\over{2}}S_{\theta2}S_{r1}\right)\right].\eeqa
The vertical velocity at third order is obtained by equation
(\ref{g4}) at $O(|\psi|^3)$ \beqa
g_{43}^\prime(\phi)&=&4g_{41}f_{42}(\phi)+g_{41}f_{31}(\phi)
g_{43}^\prime(\phi)\cos\phi\eeqa
it yields a solution as\beqa
g_{43}(\phi)&=&-2S_{\theta2}{\tilde
C}_{\theta1}-{{1}\over{4}}S_{r1}{\tilde C}_{\theta1}. \eeqa\\
\subsection{Evaluation of the coefficients}
With respect to the functions $f_1...f_6 ,g_4$ and $g_5$, we find
truncated Taylor series for the coefficients $Q^\prime_1$ and
$Q^\prime_4$ as:\beqa
Q^\prime_1&=&Q^\prime_{10}+|\psi|^2Q^\prime_{12}+O(|\psi|^3),\\
Q^\prime_4&=&Q^\prime_{40}+|\psi|Q^\prime_{41}+|\psi|^2Q^\prime_{42}+O(|\psi|^3).
\eeqa
At zeroth order, we obtain \beqa
Q^\prime_{10}={\textstyle{{1}\over{2}}}(\tilde\kappa^2-4)\beta,\eeqa
and \beqa
Q^\prime_{40}&=&-{\rm i}\beta\big\langle{\rm e}^{{\rm
i}\phi}(f_{31}(\phi)+\sin\phi)\big\rangle\nonumber\\
&=&{\textstyle{{1}\over{2}}}\beta(1-{\rm i}Z_{r1})\nonumber\\
&=&{{1}\over{2}}\beta{{2-\tilde\kappa^2}\over{1-\tilde\kappa^2}}.
\eeqa
At first order, \beqa
Q^\prime_{41}&=&-{\rm i}\beta\big\langle{\rm e}^{{\rm
i}\phi}f_{32}(\phi)\big\rangle\nonumber\\
&=&{{-{\rm i}\beta}\over{2}}Z_{r2}\nonumber\\
&=&-3\beta\tilde C_{\theta1}{{(\beta_{\rm
b}+{\textstyle{{1}\over{3}}}\beta)(\tilde\kappa^2-1)+\beta
\tilde\kappa^2}\over{(\tilde\kappa^2-1)^2}}.\eeqa
At second order,\beqa
Q^\prime_{12}&=&{1\over{2}}\beta
Z_{\phi1}\nonumber\\
&=&{\beta\over{4}}{\tilde\kappa^2\over{(1-\tilde\kappa^2)}}, \eeqa
and \beqa
Q^\prime_{42}&=&{\rm i}\beta\big\langle{\rm e}^{{\rm
i}\phi}(f_{42}(\phi)\cos\phi+f_{31}(\phi)\cos^2\phi-f_{33}(\phi))\big\rangle\nonumber\\
&=&-{{\rm i}\beta\over{2}}Z_{r3}\nonumber\\
&=&{\beta\over{8(\tilde\kappa^2-1)^3(\Gamma-3)}}\{4\Gamma(\tilde\kappa^2-1)-3(\Gamma+1)
(\tilde\kappa^2-1)-6-6\tilde\kappa^2(\tilde\kappa^2-1)\nonumber\\
&&\qquad+144(\Gamma-3)\beta\tilde
C_{\theta1}^2[\tilde\kappa^2(\tilde\kappa^2-1)(\beta_{\rm
b}+{\textstyle{{1}\over{3}}}\beta)-\beta\tilde\kappa^2]\}. \eeqa

From the coefficients $Q_1$ and $Q_4$, truncated Taylor series
were found only for the latter, \beqa
Q_4&=&Q_{40}+|\psi|Q_{41}+|\psi|^2Q_{42}+O(|\psi|^3).\eeqa
At zeroth order, \beqa
Q_{40}&=&\big\langle{\rm e}^{{\rm
i}\phi}f_{31}(\phi)\big\rangle\nonumber\\
&=&{1\over{2}}Z_{r1}\nonumber\\
&=&{{\rm i}\over{2(1-\tilde\kappa^2)}}.\eeqa
At first order,\beqa
Q_{41}&=&\big\langle{\rm e}^{{\rm
i}\phi}(f_{32}(\phi)+3{\rm i}\beta g_{41}\cos\phi)\big\rangle\nonumber\\
&=&{{\rm i}\over{2}}S_{r2}+{3\over{2}}{\rm i}\beta\tilde
C_{\theta1}\nonumber\\
&=&{3{\rm i}\over{2}}\tilde C_{\theta1}{{3(\beta_{\rm
b}+{\textstyle{{1}\over{3}}}\beta)(\tilde\kappa^2-1)^2+\beta\tilde\kappa^2+\beta(\tilde\kappa^2-1)^3}\over{(\tilde\kappa^2-1)^3}}.\eeqa\\
At second order, \beqa
Q_{42}&=&\big\langle{\rm e}^{{\rm
i}\phi}(f_{33}(\phi)-{\rm i}f_{31}(\phi)f_{42}(\phi)-{\rm i}f_{31}^2(\phi)\cos\phi)\big\rangle\nonumber\\
&=&{1\over{2}}Z_{r3}\nonumber\\
&=&-{\rm i}\beta Q^\prime_{42}.
\eeqa\\
\section{Evaluation of the general form of the angular momentum equation}
We wish to drive the general form of the angular momentum
equation. At first we adopt the equations from OG. \footnote{They
can be extracted from the Set B by integration. Note that, the
operation $\langle.\rangle$ stands for azimuthally averaged
quantities. The range of the integrations with respect to $\phi$
and $\zeta$ are from 0 to 2$\pi$ and $-\infty$ to $\infty$,
respectively.}\beqa
\lefteqn{\Sigma\bar v_r{\partial\over{\partial
r}}(r^2\Omega)=}&\nonumber\\
&&\int\big\langle{{1}\over{r^2}}\partial_r\left
[\mu_0r^4\Omega^\prime-\rho_0r^3v_{r1}(v_{\phi1}+rv_{r1}\gamma^\prime
\cos\beta_E)\right.\nonumber\\
&&\left.+\mu_0r^3(\beta^\prime_E\cos\phi+\gamma^\prime\sin\beta_E\sin\phi)
\partial_\zeta(v_{\phi1}+rv_{r1}\gamma^\prime\cos\beta_E)\right]\big\rangle\,r\,{\rm d}\zeta\nonumber\\
&&-\int\big\langle\rho_0\left[v_{\theta1}+rv_{r1}(\beta^\prime_E\cos\phi+\gamma^\prime
\sin\beta_E\sin\phi)\right]\left[r\Omega\zeta+rv_{r1}(\beta^\prime_E\sin\phi-
\gamma^\prime\sin\beta_E\cos\phi)\right]\nonumber\\
&&-\mu_0(\beta^\prime_E\sin\phi-\gamma^\prime\sin\beta_E\cos\phi)\partial_\zeta
v_{r1}\nonumber\\
&&+\mu_0r(\beta^\prime_E\cos\phi+\gamma^\prime\sin\beta_E\sin\phi)
(\beta^\prime_E\sin\phi-\gamma^\prime\sin\beta_E\cos\phi)\partial_\zeta
\left[v_{\theta1}\right.\nonumber\\
&&\left.+rv_{r1}(\beta^\prime_E\cos\phi+\gamma^\prime
\sin\beta_E\sin\phi)\right]-\mu_0r^2\Omega(\beta^\prime_E\sin\phi-\gamma^\prime\sin\beta_E\cos\phi)^2
\big\rangle\,r\,{\rm d}\zeta.\label{I}\nonumber\\
\eeqa
and\beqa
\lefteqn{\Sigma r^2\Omega(\dot\beta_E+\bar v_r\beta^\prime_E)+{\rm
i}\Sigma r^2\Omega(\dot\gamma+\bar
v_r\gamma^\prime)\sin\beta_E=}&\nonumber\\
&&\int{{1}\over{r^2}}(\partial_r+{\rm
i}\gamma^\prime\cos\beta_E)\big\langle{\rm e}^{{\rm
i}\phi}\left\{\rho_0r^4\Omega\zeta v_{r1}-{\rm
i}\rho_0r^3v_{r1}\right.\nonumber\\
&&\left.\times\left[v_{\theta1}+rv_{r1}(\beta^\prime_E\cos\phi+
\gamma^\prime\sin\beta_E\sin\phi)\right]\right.\nonumber\\
&&\left.+{\rm
i}\mu_0r^3(\beta^\prime_E\cos\phi+\gamma^\prime\sin\beta_E\sin\phi)
\partial_\zeta\left[v_{\theta1}+rv_{r1}(\beta^\prime_E\cos\phi+
\gamma^\prime\sin\beta_E\sin\phi)\right]\right.\nonumber\\
&&\left.-{\rm i}\mu_0r^2\partial_\zeta v_{r1}-{\rm
i}\mu_0r^4\Omega(\beta^\prime_E\sin\phi-\gamma^\prime\sin\beta_E\cos\phi)
\right\}\big\rangle\,r\,{\rm d}\zeta+\int\big\langle{\rm e}^{{\rm
i}\phi}\left\{-\rho_0(r^2\Omega)^\prime\zeta
v_{r1}\right.\nonumber\\
&&\left.-\rho_0(v_{\phi1}+rv_{r1}\gamma^\prime\cos\beta_E)
\left[v_{\theta1}+rv_{r1}(\beta^\prime_E\cos\phi+\gamma^\prime
\sin\beta_E\sin\phi)\right]\right.\nonumber\\
&&\left.+{\rm i}\rho_0r\Omega\zeta(v_{\phi1}
+rv_{r1}\gamma^\prime\cos\beta_E)
\right.\nonumber\\
&&\left.+{\rm
i}\rho_0rv_{r1}(v_{\phi1}+rv_{r1}\gamma^\prime\cos\beta_E)
(\beta^\prime_E\sin\phi-\gamma^\prime\sin\beta_E\cos\phi)\right.\nonumber\\
&&\left.-
{{\mu_0}\over{r}}\partial_\zeta(v_{\phi1}+rv_{r1}\gamma^\prime
\cos\beta_E)-{\rm
i}\mu_0r(\beta^\prime_E\sin\phi-\gamma^\prime\sin\beta_E\cos\phi)\right.\nonumber\\
&&\left.\times
\left[r\Omega^\prime+(\beta^\prime_E\cos\phi+\gamma^\prime\sin\beta_E
\sin\phi)\partial_\zeta(v_{\phi1}+rv_{r1}\gamma^\prime\cos\beta_E)
\right]\right\}\big\rangle\,r\,{\rm d}\zeta.\nonumber\\
\label{II}\eeqa
where \beqa
\Sigma\bar
v_r&=&\int\big\langle\rho_0v_{r2}+\rho_1v_{r1}\big\rangle\,r\,{\rm
d}\zeta,\qquad\Sigma=\int\rho_0\,r\,{\rm d}\zeta\eeqa
here $\bar v_r(r,t)$ and $\Sigma(r,t)$ are the mean radial
velocity and the surface density, respectively.
To proceed, we define\beqa
f_6(\phi-\chi)=\tilde{\cal I}/{\cal I}\label{If6}\eeqa
where $\tilde{\cal I}(r,t)$ is the second vertical moment of the
density being defined through\beqa
\tilde{\cal I}&=&\int\rho_0\zeta^2\,r^3\,{\rm d}\zeta\eeqa
so that ${\cal I}$ is its azimuthal average, i.e. ${\cal
I}=<\tilde{\cal I}>$ and also the definition of $f_6$ requires
$<f_6>=1$. Therefore, by Substituting the relations
(\ref{h0})$\dots$(\ref{zeta2}), (\ref{find1}), (\ref{find2}) and
three defined relations (\ref{mu}), (\ref{mub}) and (\ref{If6})
into the equations (\ref{I}) and (\ref{II}), the first integral in
(\ref{I}) becomes\beqa
I_1&=&{1\over{r}}\partial_
r[\big\langle{{\tilde\kappa^2-4}\over{2}}r^4\Omega^2\beta\big\rangle\Sigma-{\big\langle
r^2\Omega^2f_3f_5f_6\big\rangle \cal I}+\big\langle
r^4\Omega^2\beta
f_5|\psi|\cos\phi\big\rangle\Sigma\nonumber\\
&&+3\big\langle r^2\Omega^2\beta
g_5f_6|\psi|cos\phi\big\rangle\cal
I],\label{I1}\nonumber\\
\eeqa
and the second integral in (\ref{I}) becomes\beqa
I_2&=&{-\big\langle \Omega^2f_4f_6\big\rangle\cal
I}-{\big\langle\Omega^2f_3f_4f_6|\psi|\sin\phi\big\rangle\cal
I}-\big\langle r^2\Omega^2\beta
f_3|\psi|\sin\phi\big\rangle\Sigma\nonumber\\
&&+\big\langle r^2\Omega^2\beta
f_4|\psi|^2\cos\phi\sin\phi\big\rangle\Sigma+{3\big\langle\Omega^2\beta
g_4f_6|\psi|^2\cos\phi\sin\phi\big\rangle\cal
I}\nonumber\\
&&+\big\langle r^2\Omega^2\beta
f_3|\psi|^3\cos^2\phi\sin\phi\big\rangle\Sigma-\big\langle
r^2\Omega^2\beta|\psi|^2\sin^2\phi\big\rangle\Sigma\nonumber\\
&&-{\big\langle\Omega^2f_3f_6|\psi|\cos\phi\big\rangle\cal
I}-{\big\langle\Omega^2f_3^2f_6|\psi|^2\sin\phi\cos\phi\big\rangle\cal
I}.\label{I2}\eeqa
similarly, the first integral in (\ref{II})[Note that, for any
function F, we can write, $\big\langle{\rm e}^{{\rm
i}\phi}F(\phi-\chi)\big\rangle=\big\langle{\rm e}^{{\rm
i}(\phi+\chi)}F(\phi)\big\rangle={\psi\over{|\psi|}}\big\langle{\rm
e}^{{\rm i}\phi}F(\phi)\big\rangle$] becomes\beqa
I^\prime_1&=&{{1}\over{r}}(\partial_r+{\rm
i}\gamma^\prime\cos\beta_E)[{\big\langle{\rm e}^{{\rm
i}\phi}r^2\Omega^2(1-{\rm i}f_4-{\rm i}f_3|\psi|\cos\phi)f_3f_6
\big\rangle\cal I}{\psi\over{|\psi|}}\nonumber\\
&&+{\rm i}\big\langle{\rm e}^{{\rm i}\phi}r^4\Omega^2\beta
f_4|\psi|\cos\phi\big\rangle \Sigma{\psi\over{|\psi|}}+ 3{\rm
i}{\big\langle{\rm e}^{{\rm i}\phi}r^2\Omega^2\beta
g_4f_6|\psi|\cos\phi\big\rangle\cal
I}{\psi\over{|\psi|}}\nonumber\\
&&+{\rm i}\big\langle{\rm e}^{{\rm i}\phi}r^4\Omega^2\beta
f_3|\psi|^2\cos^2\phi\big\rangle\Sigma {\psi\over{|\psi|}}-{\rm
i}\big\langle{\rm e}^{{\rm
i}\phi}r^4\Omega^2\beta(f_3+|\psi|\sin\phi)\big\rangle\Sigma
{\psi\over{|\psi|}}],\label{I'1}\eeqa
and finally for the second integral in (\ref{II}),we have\beqa
I^\prime_2&=&-{\big\langle{\rm e}^{{\rm
i}\phi}{\tilde\kappa^2\over{2}}\Omega^2f_3f_6\big\rangle\cal
I}{\psi\over{|\psi|}}-{\big\langle{\rm e}^{{\rm
i}\phi}\Omega^2(f_4f_5+f_3f_5|\psi|\cos\phi)f_6\big\rangle\cal
I}{\psi\over{|\psi|}}+{\rm i}{\big\langle{\rm e}^{{\rm
i}\phi}\Omega^2f_5f_6\big\rangle\cal
I}{\psi\over{|\psi|}}\nonumber\\
&&\qquad+{\rm i}{\big\langle{\rm e}^{{\rm
i}\phi}\Omega^2f_3f_5f_6|\psi|\sin\phi\big\rangle\cal
I}{\psi\over{|\psi|}}-\big\langle{\rm e}^{{\rm
i}\phi}r^2\Omega^2\beta f_5\big\rangle\Sigma
{\psi\over{|\psi|}}-3{\big\langle{\rm e}^{{\rm
i}\phi}\Omega^2\beta f_6g_5\big\rangle\cal
I}{\psi\over{|\psi|}}\nonumber\\
&&\qquad-{\rm i}\big\langle{\rm e}^{{\rm
i}\phi}{{\tilde\kappa^2-4}\over{2}}r^2\Omega^2\beta|\psi|
\sin\phi\big\rangle\Sigma{\psi\over{|\psi|}}-{\rm
i}\big\langle{\rm e}^{{\rm
i}\phi}r^2\Omega^2f_5\beta|\psi|^2\sin\phi\cos\phi\big\rangle
\Sigma{\psi\over{|\psi|}}\nonumber\\
&&\qquad-3{\rm i}{\big\langle{\rm e}^{{\rm
i}\phi}\Omega^2g_5f_6\beta|\psi|^2\sin\phi\cos\phi\big\rangle\cal
I}{\psi\over{|\psi|}}.\label{I'2}\eeqa
in which $f_n$ stands for $f_n(\phi)$.

Now, we attempt to arrange them in terms of coefficients $Q_1,
Q_2, Q^\prime_1, Q^\prime_2$ and $Q_4=Q_2+{\rm i}Q_3,
Q^\prime_4=Q^\prime_2+{\rm i}Q^\prime_3$. Then, we have\beqa
I_1&=&{1\over{r}}\partial_r[Q_1r^2\Omega^2{\cal I}+Q^\prime_1
r^4\Omega^2\Sigma],\label{I1new}\\
I_2&=&-{Q_2\Omega^2\cal
I}|\psi|^2-Q^\prime_2r^2\Omega^2\Sigma|\psi|^2.\label{I2new}\eeqa
and also\beqa
I^\prime_1&=&{{1}\over{r}}(\partial_r+{\rm
i}\gamma^\prime\cos\beta_E)(Q_4r^2\Omega^2{\cal
I}\psi+Q^\prime_4r^4\Omega^2\Sigma\psi),\label{I'1new}\\
I^\prime_2&=&Q_1{\cal I}\Omega^2\psi+Q^\prime_1\Sigma
r^2\Omega^2\psi .\label{I'2new}\eeqa

Therefore, it is possible to get the coefficients $Q_n$ and
$Q^\prime_n \{i.e.,n=1,\dots,4\}$ in terms of $f$ and $g$. In the
next step, we may write the following combinations,\beqa
\Sigma\bar
v_r{\partial\over{\partial r}}(r^2\Omega)&=&I_1+I_2,\nonumber\\
&=&{{1}\over{r}}{{\partial}\over{\partial r}}\left(Q_1{\cal I}
r^2\Omega^2\right)-Q_2{\cal I}
r^2\Omega^2\left|{{\partial\ell}\over{\partial r}}\right|^2
+{{1}\over{r}}{{\partial}\over{\partial r}}\left(Q^\prime_1\Sigma
r^4\Omega^2\right)\nonumber\\
&&-Q^\prime_2\Sigma r^2\Omega^2\left|{{\partial\ell}\over{\partial
r}}\right|^2,\label{I12}\nonumber\\
\eeqa\beqa
\Sigma r^2\Omega\left({{\partial\ell}\over{\partial t}}+\bar
v_r{{\partial\ell}\over{\partial
r}}\right)&=&I^\prime_1+I^\prime_2,\nonumber\\
&=&Q_1{\cal I} r\Omega^2{{\partial\ell}\over{\partial
r}}+{{1}\over{r}}{{\partial}\over{\partial r}}\left(Q_2{\cal I}
r^3\Omega^2{{\partial\ell}\over{\partial r}}\right)+Q_2{\cal I}
r^2\Omega^2\left|{{\partial\ell}\over{\partial
r}}\right|^2{\ell}\nonumber\\
&&+{{1}\over{r}}{{\partial}\over{\partial r}}\left(Q_3{\cal I}
r^3\Omega^2{\ell}\times{{\partial\ell}\over{\partial
r}}\right)\nonumber\\
&& +Q^\prime_1\Sigma r^3\Omega^2{{\partial\ell}\over{\partial
r}}+{{1}\over{r}}{{\partial}\over{\partial
r}}\left(Q^\prime_2\Sigma r^5\Omega^2{{\partial\ell}\over{\partial
r}}\right)+Q^\prime_2\Sigma
r^4\Omega^2\left|{{\partial\ell}\over{\partial
r}}\right|^2{\ell}\nonumber\\
&&+{{1}\over{r}}{{\partial}\over{\partial r}}\left(Q^\prime_3
r^5\Omega^2{ \ell}\times{{\partial\ell}\over{\partial
r}}\right).\nonumber\\
\label{I'12}\eeqa
where we used relations (\ref{|psi|}), (\ref{psi}). Also it has
been considered that the disc matter will lie close to
$\theta=\pi/2$. So we could write $\ell$ in terms of spherical
polar unit vectors $(\bf {e_r,e_\theta,e_\phi})$,\beqa
{\ell}&=&-{\bf e}_\theta\big|_{\theta=\pi/2}\eeqa
hence\beqa
{{\partial\ell}\over{\partial t}}&=&-{\bf
e}_r(\dot\beta_E\cos\phi+\dot\gamma\sin\beta_E\sin\phi)+{\bf
e}_\phi[\dot\beta_E\sin\phi-\dot\gamma\sin\beta_E\sin\theta\cos\phi],\eeqa\beqa
{{\partial\ell}\over{\partial r}}&=&-{\bf
e}_r(\beta^\prime_E\cos\phi+\gamma^\prime\sin\beta_E\sin\phi)+{\bf
e}_\phi[\beta^\prime_E\sin\phi-\gamma^\prime\sin\beta_E\sin\theta\cos\phi].\eeqa

To compare the present work with Pringle's work (1992), we present
the following equations (Pringle,1992),\beqa
{{\partial\Sigma}\over{\partial t}}+{{1}\over{r}}{{\partial}\over
{\partial r}}(r\Sigma\bar v_r)=0,\label{sigma} \eeqa
for the surface density $\Sigma(r,t)$, and \beqa
{{\partial}\over{\partial t}}(\Sigma
r^2\Omega{\ell})+{{1}\over{r}}{{\partial}\over{\partial
r}}(\Sigma\bar
v_rr^3\Omega{\ell})={{1}\over{r}}{{\partial}\over{\partial
r}}\left(\nu_1\Sigma r^3{{{\rm d}\Omega}\over{{\rm
d}r}}{\ell}\right)+{{1}\over{r}}{{\partial}\over{\partial
r}}\left({\textstyle{{1}\over{2}}}\nu_2\Sigma
r^3\Omega{{\partial{\ell}}\over{\partial r}}\right),\label{L1}
\eeqa
for the angular momentum $L=\Sigma R^2\Omega\ell$, in the absence
of external torques. Here $\nu_1$ and $\nu_2$ are the viscosity
corresponding to the azimuthal and vertical shears, respectively.
From these equations, one can derive\beqa
\Sigma\bar v_r{\partial\over{\partial
r}}(r^2\Omega)={{1}\over{r}}{{\partial}\over{\partial
r}}\left(\nu_1\Sigma r^3{{{\rm d}\Omega}\over{{\rm
d}r}}\right)-{\textstyle{{1}\over{2}}}\nu_2\Sigma
r^2\Omega\left|{{\partial{\ell}}\over{\partial
r}}\right|^2,\label{L2} \eeqa
for the component of angular momentum parallel to ${\ell}$, and
\beqa
\Sigma r^2\Omega\left[{{\partial{\ell}}\over{\partial
t}}+\left(\bar v_r-\nu_1{{{\rm d}\ln\Omega}\over{{\rm
d}r}}\right){{\partial{\ell}}\over{\partial
r}}\right]={{1}\over{r}}{{\partial}\over{\partial
r}}\left({\textstyle{{1}\over{2}}}\nu_2\Sigma
r^3\Omega{{\partial{\ell}}\over{\partial
r}}\right)+{\textstyle{{1}\over{2}}}\nu_2\Sigma
r^2\Omega\left|{{\partial{\ell}}\over{\partial
r}}\right|^2{\ell},\label{L3} \eeqa
for the tilt vector.\\
Comparing equations (\ref{I12}), (\ref{I'12}) (present work) and
(\ref{L2}), (\ref{L3}) (Pringle, 1992), equations (\ref{I12}) and
(\ref{I'12}) are defined for the component of angular momentum
parallel to $\ell$ and the tilt vector,respectively.

We may simplify the equations (\ref{I12}), (\ref{I'12}) and
(\ref{sigma}) to find the general form of the angular momentum
equation as follows\beqa
\lefteqn{{{\partial}\over{\partial t}}\left(\Sigma
r^2\Omega\ell\right)+{{1}\over{r}}{{\partial}\over{\partial
r}}\left(\Sigma\bar
v_rr^3\Omega\ell\right)=}&\nonumber\\
&&{{1}\over{r}}{{\partial}\over{\partial r}}\left(Q_1{\cal I}
r^2\Omega^2\ell \right)+{{1}\over{r}}{{\partial}\over{\partial
r}}\left(Q_2{\cal I} r^3\Omega^2{{\partial\ell}\over{\partial
r}}\right) +{{1}\over{r}}{{\partial}\over{\partial r}}(Q_3{\cal I}
r^3\Omega^2\ell\times{{\partial\ell}\over{\partial r}})\nonumber\\
&&+{{1}\over{r}}{{\partial}\over{\partial
r}}\left(Q^\prime_1\Sigma r^4\Omega^2\ell
\right)+{{1}\over{r}}{{\partial}\over{\partial
r}}\left(Q^\prime_2\Sigma r^5\Omega^2{{\partial
\ell}\over{\partial r}}
\right)+{{1}\over{r}}{{\partial}\over{\partial
r}}\left(Q^\prime_3\Sigma r^5\Omega^2\ell\times{{\partial
\ell}\over{\partial r}}\right).\label{L4} \eeqa \label{lastpage}
\end{document}